\documentclass[twocolumn,preprintnumbers,amsmath,amssymb,showpacs,groupedaddress,prb]{revtex4}

\usepackage{amsmath,amsfonts,amssymb}
\usepackage{epsfig}

\def\e{\begin{equation}}
\def\f{\end{equation}}
\def\=#1{\overline{\overline{#1}}}
\def\_#1{{\bf #1}}
\def\^#1{{\bf {\hat #1}}}

\def\o{\omega}
\def\E{\varepsilon}
\def\ph{\varphi_w}
\def\M{\mu}
\def\.{\cdot}
\def\x{\times}

\def\l#1{\label{eq:#1}}
\def\r#1{(\ref{eq:#1})}
\def\d{\partial}
\def\D{\nabla}

\def\l#1{\label{eq:#1}}
\def\r#1{(\ref{eq:#1})}

\begin{document}

\textwidth=180mm
\textheight=240mm

\title{Radiation from elementary sources in a uniaxial wire medium}

\author{M\'{a}rio G. Silveirinha}
\email{mario.silveirinha@co.it.pt}
\author{Stanislav I. Maslovski}
\email{stas@co.it.pt}
\affiliation{
  Departamento de Engenharia Electrot\'{e}cnica\\
  Instituto de Telecomunica\c{c}\~{o}es, Universidade de Coimbra\\
  P\'{o}lo II, 3030-290 Coimbra, Portugal}

\date{\today}

\begin{abstract}
  We investigate the radiation properties of two types of elementary
  sources embedded in a uniaxial wire medium: a short dipole parallel
  to the wires and a lumped voltage source connected across a gap in a
  generic metallic wire. It is demonstrated that the radiation pattern
  of these elementary sources have quite anomalous and unusual
  properties. Specifically, the radiation pattern of a short vertical
  dipole resembles that of an isotropic radiator close to the
  effective plasma frequency of the wire medium, whereas the radiation
  from the lumped voltage generator is characterized by an infinite
  directivity and a non-diffractive far-field distribution.
\end{abstract}


\maketitle

\section{Introduction}

Wire media, generically defined as structured materials formed by
arrays of long metallic wires,~\cite{Brown_WM, Rotman_WM, Pendry_WM}
is perhaps the class of metamaterials whose effective response is
better understood. Particularly, during the last decade a vast body
of theoretical methods and analytical tools have been developed
which enable characterizing the effective electromagnetic response
of wire-based materials in different scenarios with great accuracy.
\cite{Maslovski_MOTL, Pokrovsky_2, WMPRB, Silv_MTT_3DWires,
Constantin_WM, IgorWM2D, IgorPoynt, MarioABC, Silv_Nonlocalrods,
MarioChap, MarioEVL, Luukkonen_2009, Yakovlev_2009,
Maslovski_quasistatic_PRB_2009, IgorCNT} However, a bit
surprisingly, the problem of radiation by localized external sources
embedded within wire media, has only been cursorily discussed in the
literature.\cite{IkonenWM, Lovat2006, Burgh2008a, Burgh2008b}

This gap can be explained in part because of the peculiar
electromagnetic response of wire media, which are typically
characterized by strong spatial dispersion in the long wavelength
limit,\cite{WMPRB} and this property greatly complicates the
analytical modeling. In simple terms, a medium is spatially
dispersive if the polarization vector at some generic point in space
depends not only on the macroscopic electric field, but also on the
gradient of the field, and possibly higher order derivatives.\cite{Agranovich}

\begin{figure}[t]
\centering
\epsfig{file=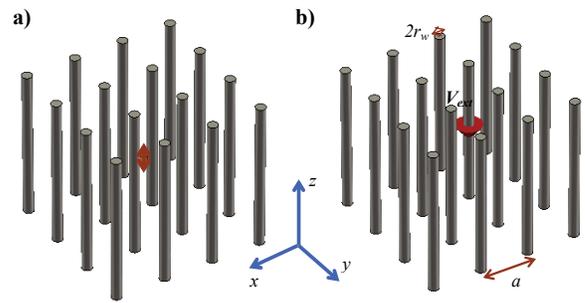, width=8.5cm}
\caption{(Color online) Uniaxial
wire medium formed by a square lattice of metallic wires oriented
along the $z$-direction. (a) Excitation based on a short vertical
dipole embedded in the wire medium. (b) Excitation based on a
discrete voltage source connected directly at the center of one of
the wires.} \label{geom}
\end{figure}

The objective of this work is to characterize the radiation
properties of elementary external localized sources placed within a
wire medium using an effective medium approach. Specifically, we are
interested in the following two scenarios: (i) a short vertical
dipole is embedded in the wire medium (Fig.~\ref{geom}a) and (ii) an
external lumped voltage source is connected across a gap in a
generic metallic wire (Fig.~\ref{geom}b). As will be detailed in
Section~\ref{SectII}, these sources are modeled in terms of the
Dirac delta function (Dirac-$\delta$).  It is, however, important to
make clear at the outset that an effective medium description of the
radiation problem is only possible if the source is localized on a
larger scale than the characteristic dimension of the metamaterial
(e.g., the lattice constant $a$, see Fig.~\ref{geom}). Hence, the
short vertical dipole considered here should be understood as some
external current spread over a region of space whose characteristic
diameter in the $xoy$-plane is larger or equal than $a$, but much
smaller than the wavelength.  Similarly, even though for the purpose
of illustration and discussion we say that in case (ii) the voltage
source is connected across the gap of a single wire, it is more
accurate to imagine such source as an array of voltage generators,
distributed over a region of space whose characteristic diameter is
larger than $a$, being each voltage generator connected across a gap
in a metallic wire lying within the mentioned region. With the
exception of the immediate vicinity of these sources, the solution
determined with our theory (based on the Dirac-$\delta$
distribution) should describe accurately the radiated fields.

One of the challenges in the characterization of the radiation by a
localized source within a wire medium is related to the calculation of
quantities such as the Poynting vector or the radiation intensity
(i.e.\ the power radiated per unit of solid angle). Indeed, in general
the usual form of the Poynting vector, ${\bf{S}} = {\bf{E}} \times
{\bf{H}}$, does not hold in case of spatially dispersive
materials.\cite{Agranovich, Costa_PRB} Moreover, there is no known
theory to determine the Poynting vector in a general spatially
dispersive material, and the only case that is actually understood and
for which closed analytical formulas are available, is when the
electromagnetic fields have a plane wave type-spatial variation.
\cite{Agranovich} In this work, we derive closed analytical formulas
that enable calculating explicitly the Poynting vector and the
electromagnetic energy density in uniaxial wire media for
\emph{arbitrary} electromagnetic field distributions. This is one of
the key results of the paper.

To do this, we rely on the theory of our earlier works,
\cite{Maslovski_quasistatic_PRB_2009, Maslovski_NJP_2011} where we
have shown that the effective medium response of the wire medium can
be modeled using a quasi-static model, based on the introduction of
two additional variables $I$ and $\varphi_w$. What is remarkable
about such a model, is that the relations between the macroscopic
electromagnetic fields and the additional variables $I$ and
$\varphi_w$ are $local$ in space. Therefore, such formalism enables
describing the unconventional electrodynamics of the wire medium
using a local approach, without requiring the definition of an
effective spatially dispersive dielectric function, which would lead
to nonlocal relations between the polarization vector and the
macroscopic electric field.\cite{WMPRB} It is important to mention
that the introduction of the additional variables is not just a
trick that simplifies the modeling of the wire medium: it is
actually full of physical significance, and elucidates about the
internal physical processes that determine the macroscopic response
of the metamaterial. Indeed, the variable $I$ can be understood as
the electric current that flows along the metallic wires
(interpolated in such a manner that it becomes a continuous function
defined in all space), whereas the variable $\varphi_w$ can be
understood as the average potential drop from a given wire to the
boundary of the respective unit cell (the potential is interpolated
in the same manner as the current). For more details, the reader is
referred to Refs.~\onlinecite{Maslovski_quasistatic_PRB_2009,
Maslovski_NJP_2011}.

This paper is organized as follows. In Section~\ref{SectII}, we
briefly review the quasi-static model of the wire medium, and
formulate the radiation problem for the two excitations of interest.
In Section~\ref{SectIII} we solve the pertinent radiation problem in
the spectral domain. First, we discuss the general case of a
stratified (along $z$) structure, and after this we analyze in
details the particular case of an unbounded uniform structure. In
Section \ref{SectUnbounded}, we show that for an unbounded uniform
structure the fields radiated by the elementary external sources can
be as well directly determined from the nonlocal dielectric function
of the metamaterial. After this, in Section~\ref{SectPoynt} we
derive a general Poynting theorem that expresses the conservation of
energy in wire media, and in Section~\ref{SectFarField} we use these
results to obtain the asymptotic form of the Poynting vector in the
far-field, as well as the directive gain, directivity, and the power
radiated by a short vertical dipole. The conclusions are drawn in
Section~\ref{SectConclusion}. In this work, we assume that in case
of time harmonic regime the time variation is of the form
$e^{j\omega t}$.

\section{Local formulation based on the introduction of additional variables \label{SectII}}

In Refs.~\onlinecite{Maslovski_quasistatic_PRB_2009,
  Maslovski_NJP_2011} it was shown that the internal physical
processes that determine the macroscopic response of a wide class of
wire media, are intrinsically related to the dynamics of the electric
current $I$ along the wires and the additional potential $\varphi_w$,
whose physical meaning was already discussed in Introduction. In
particular, it was proven that for the case of straight wires oriented
along the $z$-direction the macroscopic electromagnetic fields
satisfy
\begin{align}
\nabla  \times {\bf{E}} &=  - j\omega \mu _0 {\bf{H}},
\l{MaxEq1} \\
\nabla  \times {\bf{H}} &= {\bf{J}}_{ext} + \frac{I}{{A_{c}
}}{\bf{\hat z}} + j\omega \varepsilon _h {\bf{E}},
\l{MaxEq2}
\end{align}
where $\varepsilon_h$ is the permittivity of the host material, $A_c
= a^2$ is the area of the unit cell, and $a$ is the period of the
wire medium (Fig.~\ref{geom}). Notice that unlike in our previous
works,\cite{Maslovski_quasistatic_PRB_2009, Maslovski_NJP_2011}
here we admit the possibility of an external distributed current
source ${\bf{J}}_{ext}$. The electromagnetic fields are coupled to
the current $I$ and additional potential $\varphi_w$ via a set of
transmission line-type equations:
\begin{align}
\frac{{\partial \varphi _w }}{{\partial z}} &= -(Z_w+j\omega L)I +
E_z + V_{ext} A_c \delta \left( {x,y,z} \right),
\l{TLeq1} \\
\frac{{\partial I}}{{\partial z}} &= -j\omega C \varphi_w\,.
\l{TLeq2}
\end{align}
In the above, $C$, $L$, and $Z_w$ represent the capacitance,
inductance, and self-impedance of a wire per unit of length,
respectively, and explicit formulas for those parameters can be
found in our previous papers. The real part of $Z_w$ is related to
the ohmic loss in the metallic wires, whereas its imaginary part is
related to the kinetic inductance of the electrons in the metal. As
compared to Refs.~\onlinecite{Maslovski_quasistatic_PRB_2009,
Maslovski_NJP_2011} now we allow for an external lumped voltage
source (with amplitude $V_{ext}$) to be placed across a gap in the
wire in the central unit cell.  It is simple to check based on the
theory of Ref.~\onlinecite{Maslovski_quasistatic_PRB_2009}, that
this lumped voltage source is modeled by the term $V_{ext} A_c
\delta \left( {x,y,z} \right)$. Notice that similar to the current
and additional potential, the lumped generator is interpolated so
that it becomes a function defined over all space.

In the next section, we will determine the solution of the radiation
problems sketched in Fig.~\ref{geom}, based on the system of
Eqs.~\r{MaxEq1}--\r{TLeq2}. It should be emphasized that such a system
of equations is \emph{local} in the sense that all the relevant medium
parameters ($C$, $L$, $Z_w$, $\varepsilon_h$, and $\mu_0$) are
independent of the gradient and higher order derivatives of the
electromagnetic fields, $I$, and $\varphi_w$. This contrasts with the
usual formulation based on the effective dielectric function, which
does not require the introduction of additional variables but in which
the dielectric function depends explicitly on the wave
vector.\cite{WMPRB, Silv_Nonlocalrods} This will be further discussed
in Section~\ref{SectUnbounded}.

For future reference, we note that from Eqs.~\r{TLeq1}--\r{TLeq2} it follows
that
\begin{align} \frac{\partial }{{\partial z}}&\frac{1}{C}\frac{{\partial
I}}{{\partial z}} + \left( {\omega ^2 L - j\omega Z_w } \right)I \nonumber\\
&= -j\omega \left[ {E_z  + V_{ext} A_c \delta \left( {x,y,z} \right)}
\right]. \l{D2I}
\end{align}
In the above, it was supposed that $L$, $C$, and
$Z_w$ may depend on $z$ (but not on $x$ and $y$), which can happen
in case of a stratified wire medium (with direction of
stratification along $z$), such that either the permittivity of the
host medium or the radii of the wires varies with~$z$.

\section{The radiation problem \label{SectIII}}

Next, we derive the solution of the radiation problem in terms of
two scalar potentials. We admit that the external current density
describes a short vertical dipole, so that that ${\bf{J}}_{ext} =
j\omega p_e \delta \left( {x,y,z} \right){\bf{\hat z}}$, where $p_e$
represents the electric dipole moment. Since the Maxwell equations
are linear it is possible to solve the two radiation problems
sketched in Fig.~\ref{geom} simultaneously. This will be done in
what follows.

\subsection{Solution in terms of two scalar potentials for the general case of a stratified structure}

For generality, in this subsection, we admit that $L$, $C$, $Z_w$,
and $\varepsilon_h$ may depend on $z$, which as discussed
previously, may be useful to study problems of radiation in
stratified media. We look for a solution of~\r{MaxEq1}--\r{TLeq2} such
that the macroscopic electromagnetic fields are written in terms of a
Hertzian potential $\Phi$ so that
\begin{eqnarray}
{\bf{H}} &=& \nabla  \times \left\{ {j\omega \Phi {\bf{\hat z}}}
\right\},
\l{HertzH} \\
{\bf{E}} &=& \omega ^2 \mu _0 \Phi \,{\bf{\hat z}} + \nabla \left(
{\frac{1}{{\varepsilon _h }}\frac{{\partial \Phi }}{{\partial z}}}
\right). \l{HertzE}
\end{eqnarray}
It can be easily verified that~\r{HertzH}--\r{HertzE} satisfy, indeed,
the Maxwell equations~\r{MaxEq1}--\r{MaxEq2}, provided that
\begin{align}
\varepsilon_h \frac{\partial }{{\partial z}}\frac{1}{{\varepsilon_h
}}\frac{{\partial \Phi }}{{\partial z}} &+ \nabla _t^2 \Phi  + \omega
^2 \mu _0 \varepsilon _h \Phi  + \frac{I}{{j\omega A_{c} }} \nonumber\\
&=  - p_e\delta \left( {x,y,z} \right), \l{aux1}
\end{align}
where $\nabla _t^2  =
\frac{{\partial ^2 }}{{\partial x^2 }} + \frac{{\partial ^2
}}{{\partial y^2 }}$. Hence, substituting \r{HertzE} into \r{D2I}
and using the above result, it follows that
\begin{align}
C\frac{\partial }{{\partial z}}&\frac{1}{C}\frac{\partial
}{{\partial z}}\left( {\frac{I}{{j\omega A_{c} }}} \right) + \left(
{\omega ^2 LC - j\omega Z_w C} \right)\left(
{\frac{I}{{j\omega A_{c} }}} \right)
\nonumber\\
&=  - \frac{C}{{A_{c} \varepsilon _h }}  \left[ {\varepsilon _h E_z
+ \varepsilon _h V_{ext} A_c \delta \left( {x,y,z} \right)} \right]
\nonumber\\
&=  - \frac{C}{{A_{c} \varepsilon _h }}\left[ { - \nabla _t^2 \Phi
-p_{ef} \delta \left( {x,y,z} \right) - \frac{I}{{j\omega A_{c} }}} \right],
\l{aux2}
\end{align}
where we defined the effective dipole moment for the combined
excitations: \e p_{ef}  = p_e  - \varepsilon _h A_c V_{ext}\,. \f
Notice that $p_{ef}$ depends on both $p_e$ and $V_{ext}$, because we
allow for the simultaneous excitation of the wire medium with the
two pertinent types of elementary sources.

For convenience, let us introduce the auxiliary potential \e \psi =
\frac{1}{{k_p^2 }}\frac{I}{{j\omega A_c }},\l{defpotphi} \f where $k_p
= \sqrt{\mu_0/(LA_c)} $ is the so-called plasma wave number of the
wire medium,\cite{Maslovski_quasistatic_PRB_2009, WMPRB,
  Silv_Nonlocalrods} which may be calculated using, for example, the
approximate formula applicable to both square and hexagonal wire
lattices $k_p \approx (1/a)\sqrt{2\pi/\log[a^2/4r_w(a-r_w)]}$, where
$r_w$ is the radius of the metallic wires. Using the fact that for
straight unloaded wires $LC = \mu _0 \varepsilon _h$, it follows
that Eqs.~\r{aux1} and \r{aux2} are equivalent to
\begin{align}
&\varepsilon _h \frac{\partial }{{\partial z}}\frac{1}{{\varepsilon
_h }}\frac{{\partial \Phi }}{{\partial z}} + \nabla _t^2 \Phi  +
k_h^2 \Phi  + k_p^2 \psi
=  - p_e \delta \left( {x,y,z} \right),\\
&C\frac{\partial }{{\partial z}}\frac{1}{C}\frac{\partial }{{\partial
z}}\psi  + \left( {k_h^2 + \beta_c^2- k_p^2 } \right)\psi - \nabla_t^2 \Phi
= p_{ef} \delta\left( {x,y,z} \right),
\end{align}
where we put $k_h^2  = \omega ^2 \mu_0 \varepsilon_h$ and
${\beta_c^2 =  - j\omega Z_w C}$. Hence, to determine the solution
of our problem, we need to solve this coupled system of partial
differential equations with unknowns $\Phi$ and $\psi$.

To do this, it is most convenient to work in the Fourier domain.
Defining $\tilde \Phi $ and $\tilde \psi $ as the Fourier transform
of $\Phi$ and $\psi$ in the $xy$ plane,
respectively, so that \e\tilde \Phi = \int {\int {\Phi e^{j\left(
{k_x x + k_y y} \right)} dxdy} }, \f and $\tilde \psi $ is defined
similarly, it follows that
\begin{align}
&\varepsilon_h \frac{\partial }{{\partial z}}\frac{1}{{\varepsilon
_h }}\frac{{\partial \tilde\Phi }}{{\partial z}} +
\left(k_h^2-k_t^2\right) \tilde\Phi + k_p^2 \tilde\psi  =  - p_e
\delta
\left( z \right), \l{systFourier1}\\
&C\frac{\partial }{{\partial z}}\frac{1}{C}\frac{\partial }{{\partial
z}}\tilde\psi  + \left( {k_h^2  + \beta_c^2 - k_p^2 }
\right)\tilde\psi + k_t^2 \tilde\Phi  = p_{ef}
\delta \left( z \right),
\l{systFourier2}
\end{align}
where $k_t^2  = k_x^2  + k_y^2$. Thus, we have reduced the radiation
problem to the solution of a system of linear ordinary differential
equations.

\subsection{The case of a homogeneous medium}

From hereafter, we restrict our attention to the particular case of
a homogeneous and uniform medium, for which the structural
parameters $\varepsilon_h$, $C$, $L$, and $Z_w$ can be assumed
independent of $z$. In such a case, the system \r{systFourier1}--\r{systFourier2} can be rewritten in a compact matrix notation as follows:
\begin{align}
\frac{{\partial ^2 }}{{\partial z^2 }}\left( {\begin{array}{*{20}c}
   {\tilde \Phi }  \\
   {\tilde \psi }
\end{array}} \right) &= \left( {
\begin{array}{*{20}c}
   {k_t^2  - k_h^2 } & { - k_p^2 }  \\
   { - k_t^2 } & {k_p^2  - k_h^2  -\beta_c^2}
\end{array}} \right)\.
\left( {\begin{array}{*{20}c}
   {\tilde \Phi }  \\
   {\tilde \psi }
\end{array}} \right)\nonumber\\
&+ \,\, \delta \left( z \right)\left( {
\begin{array}{*{20}c}
   { - p_e }  \\
   {p_{ef} }
\end{array}} \right).
\l{matsyst}
\end{align}
The general solution of the homogeneous problem, when $p_e = p_{ef}
=0$, can be easily found using standard methods, and is given by
\begin{align}
\left( {\begin{array}{*{20}c}
   {\tilde \Phi }  \\
   {\tilde \psi }
\end{array}} \right) &=
\left( {C_1^ +  e^{ - \gamma _{TM} z}  +
C_1^ -  e^{ + \gamma _{TM} z} } \right)\left( {
\begin{array}{*{20}c}
   {k_p^2 }  \\
   {\gamma _h^2  - \gamma _{TM}^2 }
\end{array}} \right) \nonumber\\
&+ \left( {C_2^ +  e^{ - \gamma _{qT} z}  + C_2^ -  e^{ + \gamma
_{qT} z} } \right)\left( {
\begin{array}{*{20}c}
   {k_p^2 }  \\
   {\gamma _h^2  - \gamma _{qT}^2 }
\end{array}} \right).
\end{align}
where $\gamma_h^2 = k_t^2  - k_h^2$, ${C_i^\pm}$ with $i=1,2$ are
integration constants, and $\gamma _{qT}$ and $\gamma _{TM}$ are the
propagation constants along the $z$ direction of the so-called
quasi-transverse electromagnetic ($qT$) and transverse magnetic
($TM$) modes supported by the bulk wire medium. These parameters are
defined consistently with Refs.~\onlinecite{Silv_Nonlocalrods, WMIR}, and
satisfy
\begin{align}
\l{gammaTM}
\gamma _{TM}  = &j\left[ k_h^2  - \frac{1}{2}\left( k_p^2 + k_t^2
- \beta _c^2 + \right.\right.\nonumber\\
&\left. \left.\sqrt
{\left( {k_p^2 + k_t^2 - \beta _c^2 } \right)^2  +
4k_t^2 \beta _c^2 }
 \right) \right]^{\frac{1}{2}},
\\
\l{gammaqT}
\gamma _{qT}  = &j\left[ k_h^2  - \frac{1}{2}\left( k_p^2 + k_t^2 -
\beta _c^2 - \right.\right.\nonumber\\
&\left. \left. \sqrt
{\left( {k_p^2 + k_t^2 - \beta _c^2 } \right)^2 + 4k_t^2 \beta _c^2}
 \right) \right]^{\frac{1}{2}}.
\end{align}
In case of perfectly conducting wires, we have $Z_w=0$, and thus
$\beta_c=0$. In such a case the propagation constants of the $qT$
and $TM$ modes reduce to the well known forms, $\gamma_{qT}  =
jk_h $ and $\gamma_{TM}  = \sqrt{k_p^2  + k_t^2  - k_h^2 }$,
respectively.\cite{WMPRB}

Since the solution of \r{matsyst} is obviously an even function of
$z$, we may try a solution of the form
\begin{align}
\left( {\begin{array}{*{20}c}
   {\tilde \Phi }  \\
   {\tilde \psi }
\end{array}} \right) &=
C_{qT} e^{ - \gamma _{qT} \left| z \right|} \left( {
\begin{array}{*{20}c}
   {k_p^2 }  \\
   {\gamma _h^2  - \gamma _{qT}^2 }
\end{array}} \right) \nonumber\\
&+C_{TM} e^{ - \gamma _{TM} \left| z \right|} \left( {
\begin{array}{*{20}c}
   {k_p^2 }  \\
   {\gamma _h^2  - \gamma _{TM}^2 }
\end{array}} \right).
\l{attemptsol}
\end{align}
By direct substitution into \r{matsyst}, it is readily found that
the unknown constants $C_{qT}$ and $C_{TM}$ are required to satisfy
\begin{align}
\gamma_{qT} C_{qT} \left( {
\begin{array}{*{20}c}
   {k_p^2 }  \\
   {\gamma_h^2  - \gamma_{qT}^2 }
\end{array}} \right) &+ \gamma _{TM} C_{TM}  \left( {
\begin{array}{*{20}c}
   {k_p^2 }  \\
   {\gamma _h^2  - \gamma _{TM}^2 }
\end{array}} \right)  \nonumber\\
&= -\frac{1}{2} \left( {
\begin{array}{*{20}c}
   { - p_e }  \\
   {p_{ef} }
\end{array}} \right).
\end{align}
This yields
\begin{align}
C_{qT}  &= \frac{1}{{2\gamma _{qT} }}\frac{{ \left( {\gamma _h^2 -
\gamma _{TM}^2 } \right) p_e + k_p^2 p_{ef} }}{{\gamma _{qT}^2 -
\gamma _{TM}^2 }}\frac{1}{{k_p^2 }},
\\
C_{TM}  &= \frac{1}{{2\gamma _{TM} }}\frac{{ \left( {\gamma _h^2 -
\gamma _{qT}^2 } \right) p_e + k_p^2 p_{ef} }}{{\gamma _{TM}^2  -
\gamma _{qT}^2 }}\frac{1}{{k_p^2 }}.
\end{align}
Substituting this result into Eq. \r{attemptsol}, we finally obtain
the desired solution
\begin{widetext}
\begin{multline}
\left( {
\begin{array}{*{20}c}
   {\tilde \Phi }  \\
   {\tilde \psi }  \\
\end{array}} \right) =
\frac{1}{{2\gamma _{qT} }}\frac{{\left( {\gamma _h^2  - \gamma _{TM}^2 } \right)p_e  + k_p^2 p_{ef} }}{{\gamma _{qT}^2  - \gamma _{TM}^2 }}\,e^{ - \gamma _{qT} \left| z \right|} \left(
{\begin{array}{*{20}c}
   1  \\
   {\frac{{\gamma _h^2  - \gamma _{qT}^2 }}{{k_p^2 }}}  \\
\end{array}} \right) + \\
\frac{1}{{2\gamma _{TM} }}\frac{{\left( {\gamma _h^2  - \gamma _{qT}^2 } \right)p_e  + k_p^2 p_{ef} }}{{\gamma _{TM}^2  - \gamma _{qT}^2 }}\,e^{ - \gamma _{TM} \left| z \right|} \left(
{\begin{array}{*{20}c}
   1  \\
   {\frac{{\gamma _h^2  - \gamma _{TM}^2 }}{{k_p^2 }}}  \\
\end{array}} \right). \l{exactsol}
\end{multline}
\end{widetext}
The inverse Fourier transform of $\tilde \Phi $ is given by
\begin{align}
\Phi  &= \frac{1}{{\left( {2\pi } \right)^2 }}\int {\int {\tilde
\Phi e^{ - j\left( {k_x x + k_y y} \right)} } dk_x dk_y }\nonumber\\
&= \frac{1}{{2\pi }}\int\limits_0^{ + \infty } {\tilde \Phi \,J_0
\left( {k_t \rho } \right) k_t dk_t }, \l{inverseFourier}
\end{align}
where $J_0$ is the zero-order Bessel function of the first kind, $
\rho  = \sqrt {x^2  + y^2 }$, and in the second identity we used the
fact that $\tilde \Phi$ is a function of $k_t$. In general, this
Sommerfeld-type integral can only be evaluated using numerical
methods. Obviously, it is possible to write a similar formula for
$\psi$.

\subsection{Perfectly electric conducting wires
\label{SectPECwires}}

Let us now study what happens when to a first approximation the
metal can be modeled as a perfect electric conductor (PEC), so that $Z_w
\approx 0$. In such a situation, Eq. \r{exactsol} simplifies to
\begin{align}
\left( {\begin{array}{*{20}c}
   {\tilde \Phi }  \\
   {\tilde \psi }
\end{array}} \right) &= \frac{1}{{2\gamma _{qT} }}k_p^2 \frac{{p_e  - p_{ef} }}{{k_p^2  + k_t^2 }}\,e^{ - \gamma _{qT} \left| z \right|} \left( {
\begin{array}{*{20}c}
   1  \\
   {\frac{{k_t^2 }}{{k_p^2 }}}
\end{array}} \right)\nonumber\\
&+\frac{1}{{2\gamma _{TM} }}\frac{{k_t^2 p_e  + k_p^2 p_{ef}
}}{{k_p^2 + k_t^2 }}\,e^{ - \gamma _{TM} \left| z \right|} \left(
{\begin{array}{*{20}c}
   1  \\
   { - 1}
\end{array}} \right).
\l{solPEC}
\end{align}
We will discuss separately the two scenarios of interest. Let us
consider first that $V_{ext}=0$, so that the metamaterial is excited
solely with the short vertical dipole. In this case, $p_{ef}=p_e$,
and thus we obtain simply
\e \left( {\begin{array}{*{20}c}
   {\tilde \Phi }  \\
   {\tilde \psi }  \\
\end{array}} \right) = p_e \frac{1}{{2\gamma _{TM} }}e^{ - \gamma _{TM} \left| z \right|} \left( {\begin{array}{*{20}c}
   1  \\
   { - 1}  \\
\end{array}} \right).
\f
The corresponding inverse Fourier transforms can be evaluated
analytically in a trivial manner. This yields
\e \left(
{\begin{array}{*{20}c}
   \Phi   \\
   \psi   \\
\end{array}} \right) = p_e \frac{1}{{4\pi r}}e^{ - jk_{ef} r} \left( {\begin{array}{*{20}c}
   1  \\
   { - 1}  \\
\end{array}} \right),
\l{PhiMario} \f
with $k_{ef}  = \sqrt {k_h^2  - k_p^2 }$ and $r =
\sqrt {x^2 + y^2  + z^2 }$.

Hence, it is immediately seen that the $qT$ mode (which in the case
of a lossless metal is exactly transverse electromagnetic mode
($TEM$) with respect to the $z$-direction \cite{WMPRB}) does not
contribute to the radiation field of the short vertical dipole. This
may look surprising at first, but it is actually simple to
understand. Indeed, it is well known that the electric Green dyadic
in a periodic structure (e.g. a photonic crystal or a metamaterial)
can be written as a weighted summation of terms such as ${\bf{E}}_n
\otimes {\bf{E}}_n$ where $ {\bf{E}}_n$ stands for a generic natural
mode of the system, and~$\otimes$ represents the tensor
product.\cite{Sakoda} In particular, this immediately implies that a
TEM mode (with respect to the $z$-direction) cannot possibly
contribute to the field radiated by a short vertical dipole, because
its contribution would be proportional to ${\bf{E}}_{TEM} \left(
{{\bf{E}}_{TEM}  \cdot {\bf{\hat z}}} \right)$, whereas for a TEM
mode ${\bf{E}}_{TEM} \cdot {\bf{\hat z}} = 0$. Notice that this
discussion applies actually to the microscopic electromagnetic
fields (before homogenization on the scale of the lattice constant),
but it clearly indicates that the TEM mode cannot contribute as well
to the radiated field in the framework of a macroscopic theory,
consistent with Eq. \r{PhiMario}. It is interesting to note that in
the presence of loss the contribution of the $qT$ mode to the
radiation field does not vanish [the first addend of Eq.
\r{exactsol} does not vanish when $p_e=p_{ef}$], which is fully
consistent with the microscopic theory, because in case of loss the
electric field associated with the $qT$ mode has a small
longitudinal component (i.e.\ a component along the $z$ direction).

\begin{figure}[t] \centering
\epsfig{file=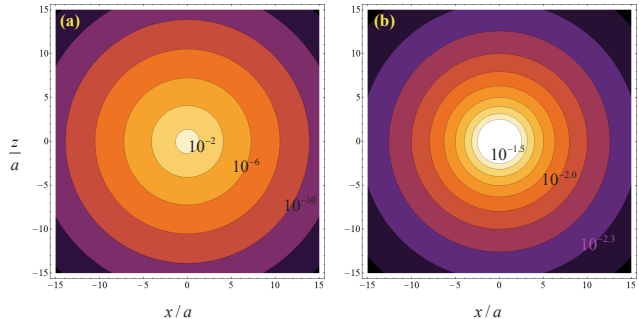, width=8.5cm} \caption{(Color online)
Contour plots of the amplitude of the potential $\Phi$ (arbitrary
logarithmic unities). (a) $\omega a/c =0.5$. (b) $\omega a/c =1.5$.
The wire medium is formed by PEC wires with $r_w = 0.01 a$ standing
in a vacuum, and is excited by a short vertical dipole.
 } \label{phidip}
\end{figure}

The result \r{PhiMario} implies two unexpected things. First,
despite the anisotropy of the wire medium, the wavefronts are
spherical surfaces! Second, the emission of radiation is
possible only above the effective plasma frequency of the
metamaterial, $ \omega _p  = k_p /\sqrt {\varepsilon _h \mu _0 }$.
The latter property is actually a direct implication of the TEM mode
not being excited, as discussed above. To illustrate variation in
space of the potential $\Phi$, we plot in Fig. \ref{phidip} the
contour plots of $\Phi$ for two different frequencies of operation.
The wire medium is formed by PEC wires with
$r_w = 0.01 a$ standing in a vacuum. The plasma wave number of the
effective medium is $k_p =1.38/a$. Thus, the example of Fig.
\ref{phidip}a corresponds to a frequency below $\omega_p$, whereas
the example of Fig. \ref{phidip}b corresponds to a frequency above
$\omega_p$. This explains that in the former case the potential is
strongly localized in the vicinity of the dipole, whereas in the
latter case the potential decays much more slowly as $1/r$.

At first glance, the result \r{PhiMario} could suggest that the
electric field radiation pattern should be similar to that of a
Hertzian dipole standing in a homogeneous isotropic plasma with a
Drude dispersion. As it will be shown in Section~\ref{SectFarField},
this is not true.

Next, we consider the case $p_e=0$, so that the wire medium is excited
by a lumped voltage source (Fig. \ref{geom}b). In this scenario, the
contribution of the TEM mode to the radiation field does not
vanish. Indeed, the inverse Fourier transform of the first term in the
right-hand side of \r{solPEC} can be readily calculated and is equal
to \e\Phi _{qT} = {- p_{ef} }\frac{1}{{2\gamma _{qT} }}e^{ - \gamma
  _{qT} \left| z \right|} \frac{{k_p^2 }}{2 \pi} K_0 \left( {k_p \rho
  } \right), \l{PhiVextTEM}\f where $K_0$ is the modified Bessel
function of the second kind. On the other hand, the auxiliary
potential~\r{defpotphi} associated with the wire current satisfies
$\psi _{qT} = - \frac{1}{{k_p^2 }}\nabla _t^2 \Phi _{qT} $.

\begin{figure}[t] \centering \epsfig{file=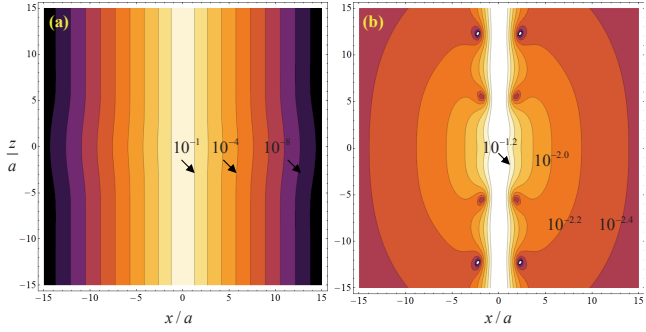,
    width=8.5cm} \caption{(Color online) Contour plots of the
    amplitude of the potential $\Phi$ (arbitrary logarithmic
    unities). (a) $\omega a/c =0.5$. (b) $\omega a/c =1.5$.  The wire
    medium is formed by PEC wires with $r_w = 0.01 a$ standing in a
    vacuum, and is excited by a lumped voltage generator. In case (b)
    an interference pattern of the TEM and TM modes is observed.}
\label{phiVext}
\end{figure}

The result~\r{PhiVextTEM} is quite remarkable, because it predicts
that the Hertzian potential, and hence the electromagnetic fields,
varies with $z$ simply as $e^{ - \gamma _{qT} \left| z \right|} = e^{
  - j k_h \left| z \right|}$, and hence the radiated field is simply
guided along $z$, without any form of decay. Moreover, $\Phi _{qT}$
is strongly localized in the vicinity of the $z$-axis, within a
spatial region whose characteristic diameter is determined by $
\lambda _p = 2\pi /k_p$. It should be mentioned that $\Phi _{qT}$ is
actually singular over the $z$-axis (it has a logarithmic
singularity). Such a singularity occurs because of the adopted
$\delta$-function model for the lumped voltage generator (also
because the macroscopic model effectively assumes infinitesimal wire
separation: $a\rightarrow 0$). The singularity disappears if one
considers a less localized model for the discrete source, e.g. if
$\delta \left( {x,y,z} \right)$ is replaced by $g\left( \rho
\right)\delta \left( z \right)$, where $g$ is some function of
$\rho$ concentrated near the origin. Even for such a source, the
electromagnetic fields are characterized by a non-diffractive
pattern. This is explained by the ``canalization'' properties of the
wire medium, which enable the transport of the near-field with no
diffraction.\cite{SWIWM, WMIR} As far as we could check, the inverse
Fourier transform of the second addend of \r{solPEC}, i.e.\ the
contribution of the TM mode when $p_e=0$, cannot be written in terms
of the standard special functions, and hence it needs to be
calculated numerically using \r{inverseFourier}. In Fig.
\ref{phiVext}, we plot the contour plots of $\Phi$, for the same
example as in Fig. \ref{phidip}. It is seen that when $\omega a/c
=0.5$ (Fig. \ref{phiVext}a), i.e.\ below the effective plasma
frequency, the fields are strongly concentrated close to the
$z$-axis, and are guided away from the source with no diffraction.
The contribution from the $TM$ mode appears to be residual. On the
other hand, above the plasma frequency (Fig. \ref{phiVext}a), there
are clearly two distinct emission channels, one associated with the
TEM mode and another with the TM mode.

\section{Nonlocal Dielectric Function Approach
\label{SectUnbounded}}

The objective of this section is to prove that the radiation problem
in an unbounded uniform structure can be as well solved using the
standard nonlocal dielectric function
formalism.\cite{Maslovski_quasistatic_PRB_2009, WMPRB,
  Silv_Nonlocalrods} The case of a stratified structure, which can be
easily handled with the theory of Section~\ref{SectIII}, is out of
reach of the nonlocal framework (it could however be handled with a
combination of mode matching and additional boundary
conditions,\cite{MarioABC, Maslovski_NJP_2011} but a detailed
discussion of it is out of the scope of this paper).

In a spatially dispersive medium, the Maxwell equations may be
written in a compact form in the space domain as follows:
\begin{align}
\nabla  \times {\bf{E}} &=  - j\omega \mu _0 {\bf{H}}, \l{Maxnonloc0}\\
\nabla  \times {\bf{H}} &= {\bf{J}}_{ext}  + j\omega
\overline{\overline \varepsilon } \left( {\omega ,j\nabla } \right)
\cdot {\bf{E}}, \l{Maxnonloc}
\end{align}
where the dyadic operator
$\overline{\overline\varepsilon}({\omega,j\nabla})$ represents the
effective dielectric function of the material. Notice that in the
space domain the effective dielectric function should be regarded as
a function of the gradient $\nabla$.  This contrasts with the
formulation of Section~\ref{SectII} where all the structural
parameters are independent of $\nabla$. It is also possible to write
the term $\overline{\overline \varepsilon } \left( {\omega ,j
\nabla} \right) \cdot {\bf{E}}$ as a spatial
convolution.\cite{Agranovich} In the spectral (Fourier) domain, in
which $j\nabla \leftrightarrow {\bf{k}}$, in the particular case of
the uniaxial wire medium formed by straight wires, the effective
dielectric function is \cite{Maslovski_quasistatic_PRB_2009,
  WMPRB, Silv_Nonlocalrods}
\begin{align}
{\=\E(\omega ,\_k)\over\E_h} &= \=I - {k_p^2\^z\^z\over k_h^2 - j\xi
k_h - k_z^2}. \l{epsok}
\end{align}
where $\xi = (Z_w/L)\sqrt{\E_h\M_0}$, and the rest of the symbols
are defined as in Section~\ref{SectII}. Notice that the effective
dielectric function depends explicitly on $k_z  \leftrightarrow
j\frac{\partial }{{\partial z}}$.

Despite the apparently complicated form of
Eqs.~\r{Maxnonloc0}--\r{Maxnonloc}, the radiation problem can be
readily solved in the spectral domain in case of an unbounded
uniform structure. Indeed, by calculating the Fourier transform of
both sides of the equations \r{Maxnonloc0}--\r{Maxnonloc} with
respect to all the space coordinates, so that $j\nabla
\leftrightarrow {\bf{k}}$, it is readily found that
\begin{align}
\l{Max1}
\_k\x\_E(\o,\_k) &= \o\M_0\_H(\o,\_k),\\
\l{Max2} \_k\x\_H(\o,\_k) &= -\o\=\E(\o,\_k)\.\_E(\o,\_k) -
\o\_P_{ext}(\o,\_k),
\end{align}
where $j\omega {\bf{P}}_{ext}\left( {\omega ,{\bf{k}}} \right) =
{\bf{J}}_{ext} \left( {\omega ,{\bf{k}}} \right)$ is the
Fourier-trans\-formed source term. After some straightforward
manipulations, we find that the Fourier transform of the electric
field is
\e {\bf{E}} = j\omega \mu _0 \left[ {\omega ^2 \mu _0
\overline{\overline \varepsilon } \left( {\omega ,{\bf{k}}} \right)
+ {\bf{kk}} - k^2 \overline{\overline {\bf{I}}} } \right]^{ - 1}
\cdot {\bf{J}}_{ext},
\f
and hence the electric field in the space
domain can be formally written as
\begin{align}
&{\bf{E}}\left( {{\bf{r}} } \right) =
\frac{{j\omega \mu _0}}{{\left( {2\pi } \right)^3 }}\x\nonumber\\
&\int {\left[ {\omega ^2 \mu_0
\overline{\overline \varepsilon } \left( {\omega ,{\bf{k}}} \right)
+ {\bf{kk}} - k^2 \overline{\overline {\bf{I}}} } \right]^{-1} }
\cdot {\bf{J}}_{ext} \left( {\bf{k}} \right)
e^{ - j{\bf{k}}\.{\bf{r}}} d^3 {\bf{k}}. \l{EsolSpec}
\end{align}
Notice that, at least \emph{a priori}, in the nonlocal dielectric
function framework we can only consider excitations based on an
external density of current (Fig.~\ref{geom}a). The characterization
of the excitation based on a lumped voltage source requires the
knowledge of internal degrees of freedom of the wire medium (e.g.,
the current along the wires and the additional potential), which are
not described by the effective medium model. Nevertheless, ahead we
will show that a lumped source $V_{ext}$ can also be modeled by a
suitable equivalent ${\bf{J}}_{ext}$.

Next, we obtain the solution of the radiation problem when
${\bf{J}}_{ext} \left( {\bf{r}} \right) = j\omega p_e {\bf{\hat
z}}\delta \left( {\bf{r}} \right)$, or equivalently when
$\_P_{ext}(\o,\_k)=p_e \^z$. Instead of attempting to calculate the
integral  \r{EsolSpec} directly, we will solve instead
\r{Max1}--\r{Max2} by introducing the Hertz potential $\_\Pi_e$. In
this manner, we write the Fourier-transformed fields as follows:
\begin{align}
\_E &= {\o^2\E_h\M_0}\_\Pi_e-\_k(\_k\.\_\Pi_e),\\
\_H &= {\o\E_h}\_k\x\_\Pi_e.
\end{align}
This form immediately satisfies \r{Max1}. From \r{Max2} we find
\e
\_k\x(\_k\x\_\Pi_e) = -{\=\E\over\E_h}\.\left[k_h^2\_\Pi_e -
  \_k(\_k\.\_\Pi_e)\right]-{p_e\^z\over\E_h}.
\f
Using~\r{epsok},
after some trivial vector algebra, we obtain
\e \l{Hertz2}
(k_h^2-k^2)\_\Pi_e = -\chi_{zz}\^z\^z\.\left[k_h^2\_\Pi_e -
  \_k(\_k\.\_\Pi_e)\right]-{p_e\^z\over\E_h}, \f where \e \l{chi}
\chi_{zz} = - {k_p^2\over k_h^2 - j\xi k_h - k_z^2}. \f Calculating
the vector product of \r{Hertz2} by $\^z$ we find that \e
\^z\x\_\Pi_e = 0, \f and therefore, $\_\Pi_e = \Pi_z\^z$. From this
and \r{Hertz2}, \e \l{Piz}
\left[k_h^2-k^2+\chi_{zz}(k_h^2-k_z^2)\right]\Pi_z =
-{p_e\over\E_h}. \f For PEC wires $\xi = 0$, and thus from~\r{chi}
we have \e (k_h^2-k_p^2-k^2)\Pi_z = -{p_e\over\E_h}, \f from which
\e \Pi_z(\o,\_k) = -{p_e\over\E_h(k_h^2-k_p^2-k^2)}, \f and, thus,
\e \Pi_z(\o,\_r) = {p_e\over\E_h}{e^{-j\sqrt{k_h^2-k_p^2}r}\over
4\pi r}, \f which is the same as \r{PhiMario}, because $\Phi =
\E_h\Pi_z$.

In the general case in which the metal has a plasmonic-type
response, $\xi \neq 0$. Introducing the notation $\beta_c^2 = -j\xi
k_h$ we obtain from \r{Piz}
\begin{align}
&\Pi_z(\o,\_k)\nonumber\\
&=-{p_e\over\E_h}
{k_h^2 + \beta_c^2 - k_z^2\over (k_h^2-k^2)(k_h^2 + \beta_c^2 - k_z^2)-k_p^2(k_h^2-k_z^2)}\nonumber\\
&=-{p_e\over\E_h}
{k_h^2 + \beta_c^2 - k_z^2\over (k_z^2+\gamma_{TM}^2)(k_z^2+\gamma_{qT}^2)},
\end{align}
where $\gamma_{TM}$ and $\gamma_{qT}$ are given by \r{gammaTM} and
\r{gammaqT}.

Calculating the inverse Fourier transform with respect to $k_z$ we find
\begin{align}
\Pi_z(\o,\_k_t,z) = {p_e\over 2\E_h}&\left({k_h^2+\beta_c^2+\gamma_{qT}^2\over\gamma_{qT}(\gamma_{qT}^2-\gamma_{TM}^2)}e^{-\gamma_{qT}|z|}\right.\nonumber\\
&+\left.{k_h^2+\beta_c^2+\gamma_{TM}^2\over\gamma_{TM}(\gamma_{TM}^2-\gamma_{qT}^2)}e^{-\gamma_{TM}|z|}\right).
\end{align}
At first glance this result looks different from~\r{exactsol}, but
one may verify that $k_h^2+\beta_c^2+\gamma_{qT}^2 =
(\gamma_h^2-\gamma_{TM}^2)+k_p^2$, and, similarly,
$k_h^2+\beta_c^2+\gamma_{TM}^2 = (\gamma_h^2-\gamma_{qT}^2)+k_p^2$.
Thus, we recover \r{exactsol} with $p_e=p_{ef}$, which corresponds
to the case $V_{ext}=0$, consistent with our assumptions in the
beginning of this section.

Surprisingly, the lumped voltage source $V_{ext}$ in
\r{TLeq1}--\r{TLeq2} can be equivalently represented {\em within the
nonlocal dielectric function model} with some distributed current
density $\_J_{ext,V}$ in the unbounded wire medium. To show this, we
consider the Fourier-transformed equations~\r{TLeq1}--\r{TLeq2} (for
simplicity we let $Z_w = 0$), from which the Fourier-transformed
current $I(\o,\_k)$ can be expressed as \e \l{ftcurrent} I(\o,\_k) =
-j\o\E_h A_c {k_p^2\over
k_h^2-k_z^2}\left[E_z(\o,\_k)+V_{ext}A_c\right].  \f

When this expression is substituted into the Fourier-transformed
equations~\r{MaxEq1}--\r{MaxEq2}, the $E_z$-proportional term
of~\r{ftcurrent} is combined with the term $j\o\E_h\_E$ which results
in the spatially dispersive permittivity~\r{epsok}, and the
$V_{ext}$-proportional term occurs as an additional external current
density \e \_J_{ext,V}(\o,\_k) = -j\o\E_hV_{ext}{k_p^2A_c\over
  k_h^2-k_z^2}\^z.  \f Therefore, applying the inverse Fourier
transform, we find that \e \l{linecur}\_J_{ext,V}(\o,\_r) =
{k_p^2A_cV_{ext}e^{-jk_h|z|}\over 2\eta_h}\delta(x,y)\^z,\f where
$\eta_h=\sqrt{\M_0/\E_h}$. Thus, a lumped voltage source inserted
into a wire of the unbounded uniaxial wire medium (with PEC wires,
$Z_w = 0$) may be equivalently represented with a line of
$z$-directed wave-like current~\r{linecur}. It is curious to note
that while the lumped voltage source excitation is localized at the
origin, the equivalent current density is distributed over the
entire $z$-axis. At first sight, this may look inconsistent with
causality. However, it is simple to verify that such a current is
just a wave emerging from the discontinuity point at $z=0$. Indeed,
if one calculates the inverse Fourier transform of \r{linecur} with
respect to time, it is found that: \e {\bf{J}}_{ext,V} \left(
{t,{\bf{r}}} \right) = k_p^2 \frac{{A_c }}{{2\eta _h }}{\tilde
V}_{ext} \left( {t - \frac{{\left| z \right|}}{{v_h }}}
\right){\bf{\hat z}}\delta \left( {x,y} \right), \f with $v_h =
1/\sqrt {\varepsilon _h \mu _0 }$ the velocity of propagation in the
host material, and ${\tilde V}_{ext}\left( t \right)$ the inverse
Fourier transform of $V_{ext}\left( \omega \right)$. The above
formula is manifestly consistent with causality, because the
excitation at a given point $z$ only depends on the excitation at
the origin with a delay ${{\left| z \right|}}/{{v_h }}$.

\section{Energy conservation in the uniaxial wire
medium and Poynting theorem \label{SectPoynt}}

In what follows, we prove that the formulation based on the
introduction of additional variables of Section~\ref{SectII},
enables formulating an energy conservation theorem and the
definition of a Poynting vector in the uniaxial wire medium.

We start with equations \r{MaxEq1}--\r{TLeq2} written in time domain.
The host permittivity $\E_h$ is assumed dispersionless and lossless,
and the wires are modeled by a self-impedance of the form $Z_w(\o) =
j\o L_{\rm kin} + R$, where the parameters $L_{\rm kin}$ and $R$ are
independent of frequency. For metallic wires with radius $r_w$
standing in air and described by the Drude model with the plasma
frequency $\o_m$ and the collision frequency $\Gamma$, these
parameters are $L_{\rm
  kin}= 1/(\E_0\pi r_w^2\o_m^2)$ and $R = \Gamma/(\E_0\pi
r_w^2\o_m^2)$.

Thus, in the time domain the equations \r{MaxEq1}--\r{MaxEq2} and \r{TLeq1}--\r{TLeq2} may be
written as
\begin{align}
\l{maxp1}
\D\x\_E &= -\M_0{\d\_H\over \d t},\\
\l{maxp2}
\D\x\_H &= \E_h{\d\_E\over \d t} + {I\over A_c}\^z + \_J_{ext},
\end{align}
and
\begin{align}
\l{tlp1}
{\d\ph\over \d z} &= -(L+L_{\rm kin}){\d I\over \d t} - RI + E_z + {\cal E}_{ext},\\
\l{tlp2}
{\d I\over \d z}  &= -C{\d\ph\over \d t},
\end{align}
where ${\cal E}_{ext}$ is the effective EMF of the voltage sources
inserted into the wires, per unit length of the wires [e.g., for a
lumped source $V_{ext}$ inserted into a wire at $\_r=0$, ${\cal
  E}_{ext}=V_{ext}A_c\delta(\_r)$].

Following a standard procedure, we obtain from~\r{maxp1}--\r{maxp2}:
\e \l{exh}\D\.[\_E\x\_H] = -{\d\over\d t}\left[{\E_h \_E^2\over 2} +
{\M_0
    \_H^2\over 2}\right] -\_E\.\_J_{ext} - {E_z I\over A_c}.  \f
On the other hand, from~\r{tlp1}--\r{tlp2} we have \e {\d(\ph
I)\over\d z} = -{\d\over\d t}\left[{C \ph^2\over 2} + {L_{\rm tot}
I^2\over
    2}\right] - RI^2 + E_zI + {\cal E}_{ext} I, \f where $L_{\rm tot}
= L + L_{\rm kin}$.

Diving the last relation by $A_c$ and adding it to \r{exh} we obtain
the conservation law
\e \l{PoyntTh} \D\.\_S = -{\d W\over\d t} -
P_{\rm loss} + P_{\rm ext}, \f where
\begin{align}
\l{Poyntvec}
\_S &= \_E\x\_H + {\ph I \over A_c}\^z,\\
\l{stor}
W &= {\E_h \_E^2\over 2} + {\M_0\_H^2\over 2} +
{C \ph^2\over 2A_c} + {L_{\rm tot} I^2\over 2A_c},\\
P_{\rm loss} &= {RI^2\over A_c},\\
\l{Pext}
P_{\rm ext} &= {{\cal E}_{ext}I\over A_c}-\_E\.\_J_{ext}\,.
\end{align}

The vectorial quantity $\_S$ in \r{PoyntTh} and \r{Poyntvec} may be
understood as the Poynting vector in the uniaxial wire medium, and
$P_{\rm ext}$ as the volume density of the power transferred by the
external sources to the medium. In the absence of loss, i.e.\ when
$R=0$, the term $W$ is univocally identified with the density of
stored energy. In contrast, if loss is present, then it is generally
impossible to separate the energy storage rate from the energy loss
rate when a metamaterial is considered {\em macroscopically}.

However, if {\em the microstructure} of a metamaterial is known, the
stored energy can be found from a consistent physical model that
fully describes the processes within a unit volume of the
metamaterial. Thus, if we assume that the Drude model is such a
consistent model for the dynamics of the free electron plasma in
metals, then \r{stor} preserves the meaning of the stored energy
density even when $R > 0$. In this case, the quantity $P_{\rm loss}$
has the physical meaning of an instantaneous power loss density.

Evidently, in time-harmonic regime the time-averaged Poynting vector
is given by \e{\bf{S}}_{av} = \frac{1}{2}{\mathop{\rm Re}\nolimits}
\left\{ {{\bf{E}} \times {\bf{H}}^* + \frac{{\varphi _w I^* }}{{A_c
      }}{\bf{\hat z}}} \right\}. \l{Sav}\f It can be checked that in
the lossless case (${\mathop{\rm Re}\nolimits} \left\{ {Z_w } \right\}
= 0$), and for the case of fields with a spatial dependence of the
form $e^{ - j{\bf{k}} \cdot {\bf{r}}}$ with ${\bf{k}}$ real-valued
this reduces to the formula, \e {\bf{S}}_{av,l} =
\frac{1}{2}{\mathop{\rm Re}\nolimits} \left\{ {\left( {{\bf{E}} \times
        {\bf{H}}^* } \right)_l } \right\} - \frac{\omega
}{4}{\bf{E}}^* \cdot \frac{{\partial \overline{\overline \varepsilon }
  }}{{\partial k_l }}\left( {\omega ,{\bf{k}}} \right) \cdot
{\bf{E}},\f with $l=x,y,z$ and $\overline{\overline \varepsilon}$
defined as in Eq.~\r{epsok}, which is applicable to plane waves in
general lossless spatially dispersive media.\cite{Agranovich,
  Silv_Poynting, Costa_PRB} The application of the above formula to
wire media has been considered in several works.\cite{ABCWM3D,
  IgorPoynt}

\section{Radiation pattern in the PEC case \label{SectFarField}}

Next, we obtain the radiation pattern, directive gain, directivity,
and radiation resistance for the case of a short vertical dipole
radiating in a wire medium formed by PEC wires (Fig. \ref{geom}a).  We
do not discuss in details the case wherein the metamaterial is excited
by a lumped voltage source, because as discussed in
Section~\ref{SectPECwires} in such a scenario the radiated field is
guided along the $z$-axis with no decay. In particular, this implies
immediately that the directivity in such a configuration is infinite.

\subsection{Asymptotic form of the radiated fields}

To begin with, we obtain the asymptotic form of the field radiated
by a short vertical dipole ($V_{ext}=0$) embedded in a wire medium
formed by PEC wires when $r \to \infty $. Evidently, from the
results of Section~\ref{SectPECwires}, unless the frequency of
operation is larger than the plasma frequency of the effective
medium the radiated fields will decay exponentially away from the
source. Hence, in what follows we assume that $\omega > \omega _p =
k_p /\sqrt {\mu _0 \varepsilon _h } $, so that $k_{ef}>0$ in
Eq.~\r{PhiMario}.  Substituting \r{PhiMario} into
Eqs.~\r{HertzH}--\r{HertzE}, it can be easily checked that
\begin{align}
{\bf{H}}\,&\dot=\, -\omega k_{ef} \Phi \sin \theta \,\hat
{\bf{\varphi}},\\
{\bf{E}}\,&\dot=\, k_h^2 \frac{\Phi }{{\varepsilon _h }}\left[
{\left( {1 - \frac{{k_{ef}^2 }}{{k_h^2 }}} \right)\cos \theta
\,{\bf{\hat r}} - \sin \theta \,\hat {\bf{\theta}} } \right],
\l{EHasymp}
\end{align}
where the symbol $\dot  =$ indicates that the identities are
asymptotic ($r \to \infty $), $\Phi  = p_e \frac{1}{{4\pi r}}e^{ -
jk_{ef} r}$, and $\left( {{\bf{\hat r}},\hat \theta ,\hat \varphi }
\right)$ define an orthogonal reference system associated with the
usual spherical coordinate system  $ \left( {r,\theta ,\varphi }
\right)$. As seen, unlike what happens in an isotropic medium, the
electric far field has a radial component. The amplitude of the
electromagnetic fields varies asymptotically as $1/r$, and $E_\theta
= \eta _{ef} H_\varphi$ with $\eta _{ef}  = \omega \mu _0 /k_{ef}$.

Similarly, substituting~\r{PhiMario} into Eqs.~\r{TLeq1}--\r{TLeq2}
and~\r{defpotphi}, it is found that the asymptotic forms of the
current and additional potential are
\begin{align}
I \,&\dot=\,  - j\omega A_c k_p^2 \Phi, \\
\varphi _w \,&\dot=\, I\frac{{k_{ef} }}{{\omega C}}\cos \theta \,\dot=\, -
\frac{{j\,k_{ef} }}{{\varepsilon _h }} \cos \theta \Phi. \l{Iasymp}
\end{align}

Thus, from~\r{EHasymp} and \r{Iasymp}, we see that the time averaged
Poynting vector \r{Sav} in the far-field is
\begin{align}
{\bf{S}}_{av}  &\,\dot=\, \frac{1}{2}\omega ^3 \mu _0 \left| \Phi
\right|^2 k_{ef} \sin \theta \left( {\frac{{k_p^2 }}{{k_h^2 }}\cos
\theta
\,\hat \theta  + \sin \theta \,{\bf{\hat r}}} \right)\nonumber\\
&+\frac{1}{2}\frac{{k_{ef} }}{{\varepsilon _h }}k_p^2 \omega \cos
\theta \left| \Phi  \right|^2 {\bf{\hat z}}.
\end{align}
Straightforward calculations show that the Poynting vector only has
a radial component: \e {\bf{S}}_{av} \,\dot =\, \frac{1}{2}k_{ef}
\omega ^3 \mu _0 \left| \Phi \right|^2 \left( {\sin ^2 \theta +
\frac{{k_p^2
      }}{{k_h^2 }}\cos ^2 \theta \,} \right){\bf{\hat r}}. \f Hence, in
part surprisingly, it is seen that a short vertical dipole embedded in
a wire medium $can$ radiate energy along the direction of vibration,
i.e. along the $z$-axis! Moreover, in the limit $\omega \to \omega _p$
the radiation pattern becomes isotropic: $ {\bf{S}}_{av} \approx
\frac{1}{2}k_{ef} \omega ^3 \mu _0 \left| \Phi \right|^2 {\bf{\hat
    r}}$. Note, however, that for $\omega =\omega _p$, we have $k_{ef} =
0$ and thus the Poynting vector vanishes in the far-field. However,
slightly above $\omega _p$ the emission of radiation is certainly
possible. Notice also that in the limit where $k_p \to 0$, we recover
the far-field of a short vertical dipole embedded in a dielectric with
permittivity $\varepsilon_h$.

\subsection{The radiation intensity, directive gain, directivity, and radiation resistance}

The radiation intensity of the short vertical dipole, $U = \lim _{r
\to \infty } r^2 {\bf{S}}_{av}$, is given by \e U = \frac{{\left|
{p_e } \right|^2 }}{{32\pi ^2 }}k_{ef} \omega ^3 \mu _0 \left( {\sin
^2 \theta  + \frac{{k_p^2 }}{{k_h^2 }}\cos ^2 \,\theta } \right). \f
Hence, the power radiated by the dipole, $P_{rad}  = \int {Ud\Omega
} = 2\pi \int {U\sin \theta d\theta }$, is such that \e P_{rad}  =
\frac{{\left| {p_e } \right|^2 }}{{12\pi }}k_{ef} \omega ^3 \mu _0
\left( {1 + \frac{{k_p^2 }}{{2k_h^2 }}} \right). \f The directive
gain, $g=4 \pi U/P_{rad}$, is \e g\left( {\theta ,\varphi } \right)
= \frac{3}{{2 + k_p^2 /k_h^2 }}\left( {\sin ^2 \theta  +
\frac{{k_p^2 }}{{k_h^2 }}\cos ^2 \theta } \right).\f Since $k_h \ge
k_p$ it can be checked that the direction of maximal radiation is
$\theta=\pi/2$. The directivity of the short vertical dipole is,
thus, \e D = \frac{3}{{2 + k_p^2 /k_h^2 }},\f which therefore
increases from unity (for $k_h \approx k_p$) up to $3/2$ in the
limit $k_h \gg k_p$.

\begin{figure}[ht]
\centering
\epsfig{file=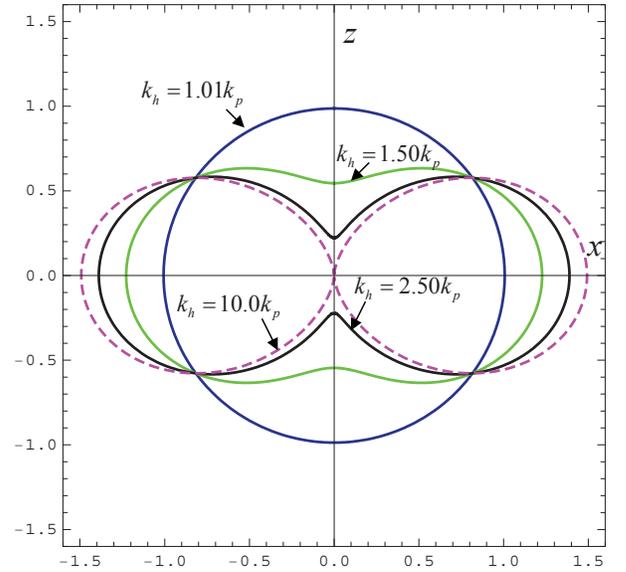, width=8.5cm} \caption{(Color
online) Polar plot of the directive gain of a short vertical dipole
embedded in the uniaxial wire medium for different frequencies of
operation ($\omega = k_h /\sqrt {\varepsilon _h \mu _0 } $).
 } \label{directivegain}
\end{figure}

In Fig.~\ref{directivegain} we show a polar plot of the directive
gain of the short vertical dipole for different frequencies of
operation, normalized to the effective plasma frequency. In
agreement with the previous discussion, it can be seen that the
radiation pattern becomes more directive for increasing values of
the frequency, and that for $\omega \approx \omega_p$ the radiator
resembles an isotropic radiator.

To conclude, we note that if the dipole is fed by a current $I_0$
and has infinitesimal height $dl$, then the corresponding dipole
moment is such that $ \omega \left| {p_e } \right| = \left| {I_0 }
\right|dl$. Thus, it follows that the radiation resistance ($R_{rad}
= 2P_{rad} /\left| {I_0 } \right|^2$) of such an elementary source
is given by
\begin{align}
R_{rad}  &= \frac{{\left( {dl} \right)^2 }}{{6\pi }}k_{ef} \omega
\mu_0 \left( {1 + \frac{{k_p^2 }}{{2k_h^2 }}} \right)\nonumber\\
&= \eta _h \frac{{\left( {dl} \right)^2 }}{{6\pi }}k_{ef} k_h \left( {1 + \frac{{k_p^2 }}{{2k_h^2 }}} \right),
\end{align}
where $\eta _h  = \sqrt {\mu _0 /\varepsilon _h }$ is the impedance
of the host material.

\section{Conclusion \label{SectConclusion}}

In this work we have studied the radiation of two types of elementary
sources embedded in a uniaxial wire medium and derived a general
energy conservation theorem. The main challenge of the radiation
problem is related to the metamaterial being spatially dispersive. We
have shown that the radiation problem can be solved by considering
either a nonlocal dielectric function framework or, alternatively, a
local model framework based on the introduction of additional
variables.  However, only the latter approach enables considering
stratified media and calculating quantities such as the Poynting
vector or the directive gain. It was shown that the emission of
radiation by a short dipole in a wire medium has several anomalous
features, such as a uniform directive gain near the effective plasma
frequency. On the other hand, the radiation by a lumped voltage
generator results in a non-diffractive beam that is localized in the
vicinity of the $z$-axis, and corresponds to an infinite directivity.

\bibliography{refs}

\end{document}